# Observation of novel gapped phases in potassium doped single layer *p*-terphenyl on Au (111)

M. Q. Ren[1], W. Chen[1], Q. Liu[1], C. Chen[1], Y. J. Qiao[2], Y. J. Chen[2], G. Zhou[2], T. Zhang[1,2,3], Y. J. Yan[1,2,3*], D. L. Feng[1,2,3*]

[1] State Key Laboratory of Surface Physics, Department of Physics, Fudan University, Shanghai 200433, China

[2] Laboratory of Advanced Materials, Fudan University, Shanghai 200433, China

[3] Collaborative Innovation Center of Advanced Microstructures, Nanjing, 210093, China

**Recently, superconductivity in potassium (K) doped *p*-terphenyl ($C_{18}H_{14}$) has been suggested by the possible observation of the Meissner effect and subsequent photoemission spectroscopy measurements, but the detailed lattice structure and more-direct evidence are still lacking. Here we report a low temperature scanning tunneling microscopy/spectroscopy (STM/STS) study on K-doped single layer *p*-terphenyl films grown on Au (111). We observe several ordered phases with different morphologies and electronic behaviors, in two of which a sharp and symmetric low-energy gap of about 11 meV opens below 50 K. In particular, the gap shows no obvious response to a magnetic field up to 11 Tesla, which would caution against superconductivity as an interpretation in previous reports of K-doped *p*-terphenyl materials. Such gapped phases are rarely (if ever) observed in single layer hydrocarbon molecular crystals. Our work also paves the way for fabricating doped two-dimensional (2D) hydrocarbon materials, which will provide a platform to search for novel emergent phenomena.**

Organic molecular materials provide an important playground in condensed matter physics, exhibiting charge density wave (CDW) phases [1], anti-ferromagnetism [1-3], Mott-insulating state [1-3] and superconductivity [2-9]. Among these, organic superconductors are of particular interest for their theoretically predicted high critical temperatures [10]. Known organic superconductors include graphite intercalation compounds [4, 5], alkali-metal doped fullerenes [2, 3], organic salts [1, 6] and alkali-metal doped aromatic hydrocarbon molecular crystals [7-9]. The record for the superconducting transition temperature (Tc) in the organic superconductors has long been 38 K for doped fullerene ($Cs_3C_{60}$ under high pressure) [2, 3] with minimal further progress obtained thereafter. Recently, reports of low-temperature diamagnetism up to 120 K in K-doped *para*-terphenyl (*p*-terphenyl or PTP) rekindled the interest for exploring possible high temperature superconductivity in doped aromatic hydrocarbon molecular crystals [11].

*p*-Terphenyl consists of three benzene rings linked by single C-C bonds in the para position. By doping K into *p*-terphenyl molecular crystals, Wang et al. observed a weak Meissner-like effect below 7.2 K, 43 K and 120 K for different doping levels [11-14]. Shortly thereafter, a gap persisting up to at least 60 K was observed in surface K-dosed *p*-terphenyl crystals by photoemission spectroscopy [15]. Taken together, these indicate possible high-temperature superconductivity in electron-doped *p*-terphenyl. However, since the reported diamagnetic signal is very low, corresponding to less than a 0.1% superconducting shielding fraction, further experiments on structural characterization and electronic properties are essential to clarify the origin of these unique phenomena. Here, *via* high vacuum deposition, we have grown well-ordered (K-doped) *p*-terphenyl films on Au (111). A clear low energy gap of about 11-13 meV is observed in two of the ordered phases, but shows no obvious response to magnetic fields up to 11 Tesla, making superconductivity an unlikely origin.

The sample preparation and experimental details are described in the Methods section. Fig. 1(a) shows a typical STM image of a cleaned Au (111) surface, which is characterized by the well-known herringbone (22 ×√3) reconstruction. Figs. 1(b) and 1(c) show STM images of a *p*-terphenyl film grown on the Au (111) substrate (labeled as PTP/Au). The *p*-terphenyl molecules lie flat on the Au (111) surface and self-assemble into a stripe-like structure. The benzene rings are distinguishable, as shown in Fig. 1(c). We note that similar structures were also reported for other surface-adsorbed phenyl molecules [16-21]. The inter-stripe and inter-molecule distances (labeled as *a* and *b* in Fig. 1(c)) are about 15.0 Å and 6.7 Å, respectively. Vacancies and domains with different stripe orientations can occasionally be observed, as indicated by the arrows and black dashed box in Fig. 1(b). Films grown at room temperature were always found to be one monolayer (ML) thick, with no terraces of a 2$^{nd}$ layer of *p*-terphenyl film observed, regardless of the growth time. (Note that the herringbone reconstruction of Au (111) is still visible through the first molecular layer, see the STM images in Figs. 1-3). Moreover, the molecules were desorbed rapidly when placed in vacuum at room temperature. This indicates rather weak bonding between the film and substrate, and even weaker interactions between the molecules. Figs. 1(d)-(f) show typical dI/dV spectra measured at 4.5 K within different energy scales on PTP/Au, as well as on bare Au (111) for comparison. There is a remarkable resemblance of the lineshape between Au (111) and PTP/Au, except for a characteristic peak about 2.7 eV above $E_F$ in Fig. 1(d). This peak may correspond to the lowest unoccupied molecular orbital (LUMO), while the reported highest occupied molecular orbital (HOMO) is about 4 eV below $E_F$ [15]. The low bias dI/dV spectra on PTP/Au display a finite but featureless density of states (DOS) near $E_F$, which is likely contributed by the Au substrate since *p*-terphenyl molecular crystals do not have low energy states [22].

Surface K doping is an effective way to introduce electron carriers into the surface layer of a sample [23, 24]. Thus we evaporate K atoms (0–3 ML coverage) onto the surface of PTP/Au to introduce electrons; here 1 ML is defined as the areal density of *p*-terphenyl molecules in single-layer PTP/Au ($8.8 \times 10^{13}$/cm$^2$). Typical topographic images and electronic properties of PTP/Au with different K coverage (Kc) are shown in Fig. s1. Several ordered phases are observed (see parts I and II of supplementary materials for details), which might be induced by different concentrations of K atoms intercalated into the *p*-terphenyl molecules. A finite but featureless DOS near $E_F$ is observed for most of these phases, while a surprising gap appears in two phases with Kc=1.45 ML, which will be discussed in detail later. Besides inducing new structural phases, K-doping prevents PTP molecules from desorbing at room temperature, indicative of charge transfer and increased bonding between PTP molecules.

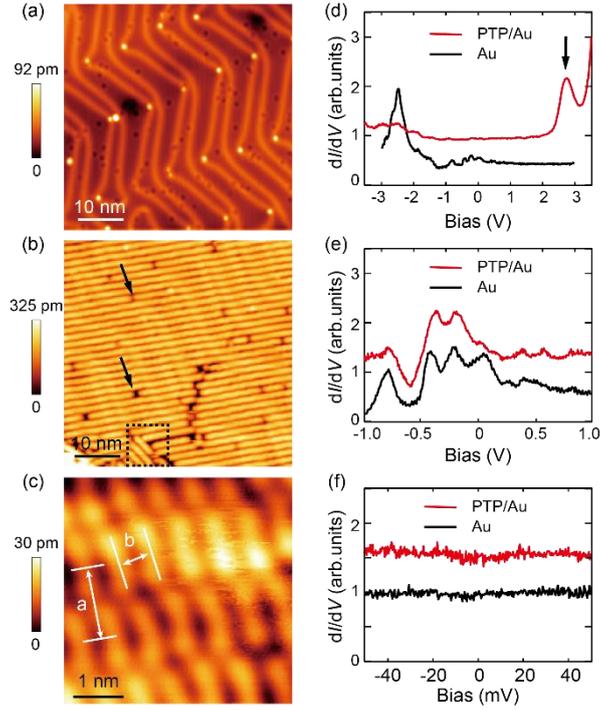

*Fig. 1. Surface topography and dI/dV spectra of PTP/Au measured at 4.5 K. (a) Typical topographic image of a clean Au (111) surface (50 × 50 nm², $V_b$=1 V, I=100 pA) – the herringbone reconstruction and point defects can be clearly seen. Typical topographic images of PTP/Au are shown in (b) (50 × 50 nm², $V_b$=3 V, I=30 pA) and (c) (5 × 5 nm², $V_b$=0.2 V, I=50 pA). Benzene rings in each p-terphenyl molecule can be distinguished. A domain with different molecular orientations is indicated by the black dashed box in (b), while the arrows point out molecular vacancies. Inter-stripe and inter-molecule distances are illustrated by white lines, and labeled as a and b. Typical dI/dV spectra of PTP/Au within different energy ranges are shown in (d): ±3.5 V ($V_b$=2 V, I=30 pA, ΔV=50 mV), (e): ±1 V ($V_b$=0.2 V, I=50 pA, ΔV=30 mV), and (f): ±50 mV ($V_b$=10 mV, I=100 pA, ΔV=1 mV). Corresponding dI/dV spectra on the bare Au (111) surface are included for comparison. The characteristic peak 2.7 eV above $E_F$ is marked by a black arrow in (d). Curves in (d)-(f) are shifted vertically for clarity.*

Next we focus on PTP/Au with Kc=1.45 ML to investigate the origin of the gapped electronic state. Four typical phases are observed in as-grown PTP/Au with Kc=1.45 ML, their detailed structural and electronic properties are shown in Fig. 2 and Fig. s2. Phases III and IV behave like a metal; phase II is gapped near $E_F$ but is rarely observed, especially after anneal treatment. Phase I is the dominant one. It is highly ordered, made up of rectangle-shaped blocks stacking alternately along the horizontal and vertical directions, as shown in Fig. 2(c). Assuming the same areal density of *p*-terphenyl molecules in pristine and doped PTP/Au, such a rectangle-shaped block contains two *p*-terphenyl molecules, as also clearly visible in the topography. Considering the detailed surface morphologies shown in Figs. 2(c) and s3, it seems that the *p*-terphenyl molecules form dimers, and about three K atoms are likely intercalated between them – Kc=1.45 ML would be 2.90 K atoms per *p*-terphenyl dimer. Such a dimer structure is consistent with the hypothetical unit proposed by Fabrizio et al. [25], which is often present in alkali-doped organics. The unit cell of phase I is indicated by the blue square box in Fig. 2(c), with a lattice constant of about 21.2 Å. Although different phases are observed, their exact structural parameters such as the orientation of the molecules, the concentration or exact positions of intercalated K atoms are difficult to ascertain directly from the topographic images. Further theoretical simulations will be necessary to fully determine the lattice structures of these different phases or identify their K-doping levels.

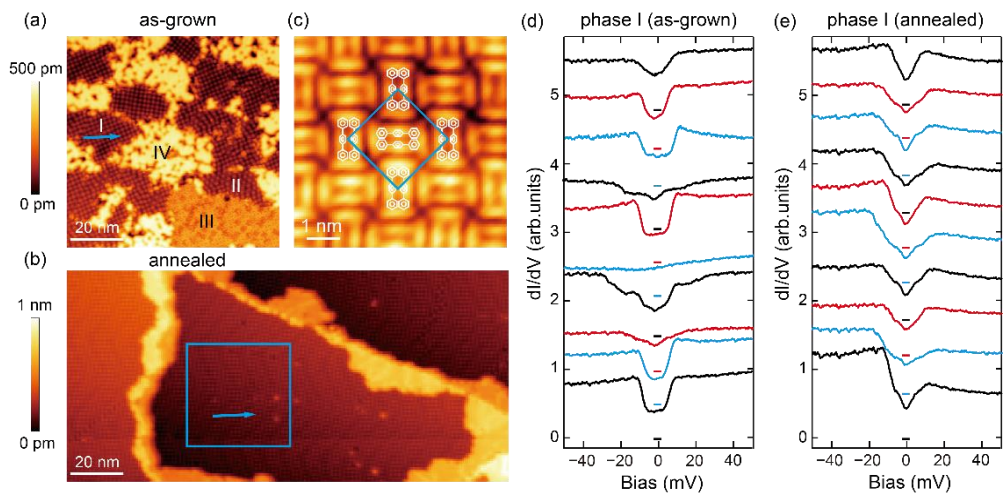

**Fig. 2.** *Typical topographic images and dI/dV spectra of PTP/Au with Kc=1.45 ML (measured at 4.5 K). (a) Typical topographic image of as-grown films (80 × 80 nm², $V_b$=1 V, I=30 pA). Four different phases are identified, their clear lattice structures are shown in (c) and Fig. s2. (b) Typical topographic image of films annealed at room temperature for 20 mins (80 × 165 nm², $V_b$=1 V, I=30 pA). The blue box in (b) indicates the area for temperature dependent experiments shown in Fig. 3. (c) Detailed lattice structure of phase I (6 × 6 nm², $V_b$=100 mV, I=100 pA). The blue square indicates the unit cell of phase I, and dimerized p-terphenyl molecules are sketched out. (d)(e) Typical dI/dV spectra of phase I in as-grown and annealed films, collected along the blue lines marked in (a) and (b) (set point: $V_b$=10 mV, I=100 pA, ΔV=1 mV). The horizontal bars in (d) and (e) indicate the zero conductance position of each curve and colour-coded with corresponding spectra.*

Two kinds of low energy gaps could be observed in the dI/dV spectra of phase I. For as-grown films without any annealing treatment, phase I is only a few to a dozen nanometers in size (see Figs. 2(a) and s4), and its electronic state has a strong spatial inhomogeneity, as shown in Fig. 2(d). A U-shaped gap is usually observed in half of these areas, with a smaller gap of about 3-4 meV occasionally observed inside. After annealing at room temperature for tens of minutes, the phase I domains grow to hundreds of nanometers in size (see Figs. 2(b)), while the gap is more homogeneous but becomes V-shaped with a double-gap structure, as seen in Fig. 2(e). Both kinds of gaps are about 10-14 meV in size, are symmetric with respect to $E_F$, and lacks obvious coherence peaks. We also note that neither gaps is fully opened, with about 50% residual DOS near $E_F$; however, it is possible that the Au substrate contributes some DOS here. The difference in lineshape of the gaps may arise from the redistribution of K atoms by annealing, which may alter the doping level or result in more consistent interactions between dimers.

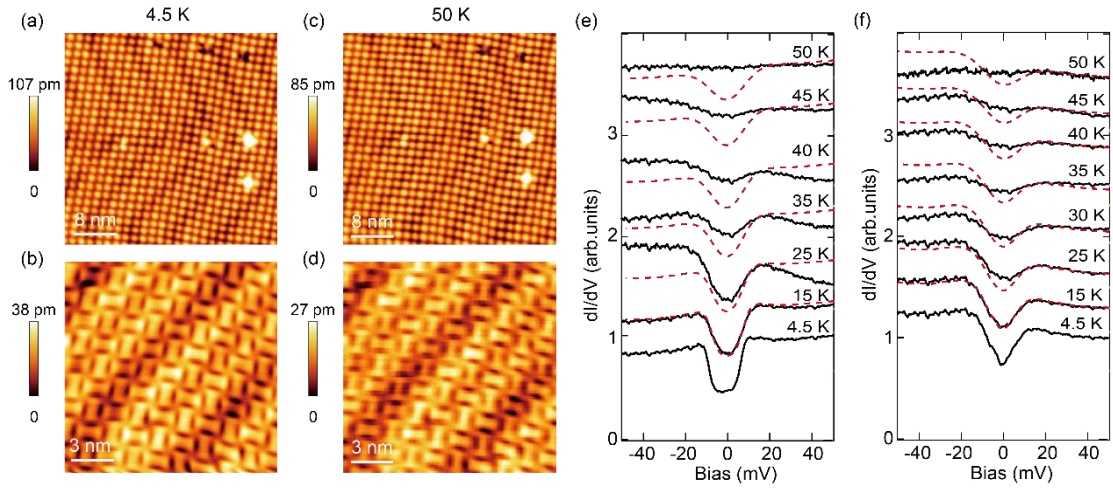

*Fig. 3. Temperature dependence of topography and dI/dV spectra of phase I observed in PTP/Au with Kc=1.45 ML. Topographic images measured at 4.5 K and 50 K are shown in (a) (b) and (c) (d), respectively. Set point: $V_b$=1 V, I=50 pA for (a) and (c) with size of 40 × 40 nm²; $V_b$=100 mV, I=50 pA for (b) and (d) with size of 15 × 15 nm². (e) Temperature dependence of a typical U-shaped gap observed in as-grown films (these spectra were collected at the same location as marked by blue dots in Fig. s4. Set point: $V_b$=10 mV, I=100 pA, ΔV=1 mV.) (f) Temperature dependence of a typical V-shaped gap observed in annealed films (these spectra were spatially averaged from 25 spectra collected within the same areas in (a)-(d). Set point: $V_b$=10 mV, I=100 pA, ΔV=1 mV.) Black curves are experimental data, red dashed curves show the effect of thermal smearing, obtained by convolving the experimental dI/dV spectrum at 4.5 K with a Fermi-Dirac function of given temperature. Curves in (e) and (f) are normalized to the value at +50 mV, and are shifted vertically for clarity.*

Temperature dependence of the topography and dI/dV spectra of phase I in PTP/Au with Kc=1.45 ML are shown in Figs. 3 and s4. The topography is temperature-independent up to 50 K, regardless of the annealing history. Figs. 3(e) and 3(f) show the temperature dependence of two kinds of gaps observed in phase I, which are collected separately in as-grown and annealed films. A U-shaped gap of about 11 meV

(at 4.5 K) gradually fills in with increasing temperature and closes at about 50 K, giving a $2\Delta/k_BT$ ratio of about 5.07. The V-shaped gap shows similar temperature dependence, and also closes at about 50 K. The usual suspects for such symmetric low energy gaps at $E_F$ include semiconducting gap, charge/spin density wave (CDW/SDW) gap, or superconducting gap. As seen in Figs. 3(e) and 3(f), with increasing temperature, the gap is suppressed much faster than thermal smearing (as indicated by the red dashed lines, see figure caption for details), eliminating the possibility of a semiconducting gap. SDW is unlikely for such a hydrocarbon system. A CDW order would manifest itself as a static spatial modulation below the transition temperature [26]. We do not observe any clear difference in the topography of phase I measured below and above the gap-closing temperature (as seen in Figs. 3(a)-(d) and Fig. s4). However since the topography may be dominated by intra- molecular states (LUMO and HOMO), weak charge modulation may still exist which cannot be readily resolved. More sensitive techniques are needed to exclude or identify CDW as the possible cause of the gaps.

We continue by measuring the magnetic field (B) dependence of this gap to determine whether it is related to superconductivity. Figs. 4(c) and 4(d) show the zero-bias conductance (ZBC) map in a 50 × 50 nm$^2$ area of phase I (shown in Fig. 4(a)) measured under H=0 T and 11 T, respectively. There is a clear phase separation due to the multi-domain structure of the doped PTP/Au film and inhomogeneous K distribution. The light green areas show metallic behavior, while the above mentioned low energy gap is observed in dark blue areas. No obvious magnetic vortices can be distinguished in the ZBC map under H=11 T, and almost all of the dI/dV spectra within the field of view are unchanged with increased H, as illustrated in Figs. 4(e) for representative V-shaped spectra measured along a line within one domain under H=0 T and 11 T, respectively. The U-shaped gap is similarly insensitive to H, as shown in Fig. 4(f), the absence of a smaller gap at some locations under H=0 T is induced by impurity effect (see Fig. s5 for more details). In a type-II superconductor, the superconducting order parameter recovers within about twice its coherence length ($\xi$) away from the vortex core (Fig. s6 illustrates the 1ML FeSe/STO case with $\xi$ of about 2-3 nm as an example) [27-29]. And in an imperfect superconductor, the magnetic vortices are easily pinned by impurities which suppress superconductivity [29], or magnetic field may leak through non-superconducting areas. The vortex spacing under an 11 T field is estimated to be about 14 nm (assuming a triangular vortex lattice), which is equal to the size of the domain with the gapped phase in our case. Thus the magnetic field may leak entirely through the light green areas without a gap, leading to the observed absence of magnetic vortices. Nevertheless, suppression of the gap may still happen near the boundary of the gapped phase, on the spatial scale of $2\xi$. No obvious field dependence was observed in dI/dV spectra at locations 0.6 nm away from metallic areas under H = 0 T and 11 T, as shown by curves 2 and 3, 13 and 14 in Figs. 4(e). If phase I is indeed superconducting, this would give an in-plane $\xi$ of less than 3 Å, not only smaller than the size of one unit cell, but significantly shorter than even a single molecule. Such a short in-plane $\xi$ has never been reported for any superconductor. Therefore, our data indicate that the observed gap is unlikely to be a superconducting gap.

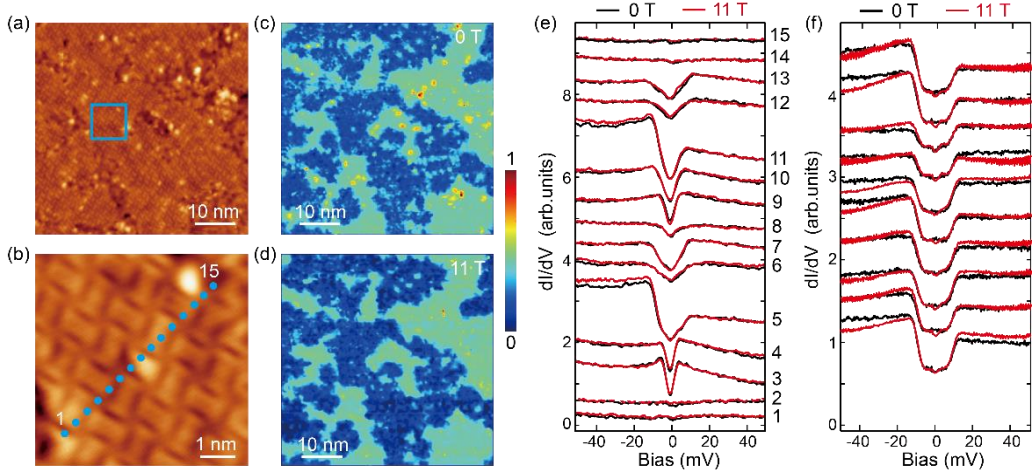

**Fig. 4**. *Magnetic field dependence of electronic state observed in phase I of PTP/Au with Kc=1.45 ML. (a) Topographic image of phase I for dI/dV mapping (size: 50 × 50 nm$^2$, set point: $V_b$=100 mV, I=10 pA). Blue square is a region within a single domain, its topographic image is shown in (b) (size: 6 × 6 nm$^2$, set point: $V_b$=100 mV, I=10 pA). (c)-(d) Zero-bias conductance (ZBC) maps (set point: $V_b$=20 mV, I=100 pA, ΔV=2 mV), taken under magnetic fields of 0 T and 11 T, respectively. Clear phase separation appears, and no magnetic vortices can be distinguished. (e) and (f) show magnetic field dependence of typical V-shaped and U-shaped gaps, respectively. Set point: $V_b$=50 mV, I=60 pA, ΔV=2 mV. Spectra 1-15 in (e) are collected at the locations indicated by blue dots in (b), with an interval between adjacent points of 0.6 nm. The spectra in (f) were taken in a small area (size: 1 × 1 nm$^2$) where U-shaped gaps exist ubiquitously. Black and red curves are measured under H=0 T and 11 T, respectively. Spectra in (e) and (f) are shifted vertically for clarity*

The gaps observed here in doped PTP/Au are symmetric and do not exhibit any coherence peaks, which resembles the behaviors of the pseudogap phase in cuprates [27]. As discussed above, the common origins, such as CDW, SDW and superconductivity, are likely not the cause for the gapped electronic states. Therefore, other candidates for the observed gap need to be considered. As in the intriguing pre-pairing interpretation for the cuprate pseudogap phase, the low energy gaps in doped PTP/Au could be due to the local pairing of electrons, and even likely in the form of bipolaron pairing in such a hydrocarbon system [30]. These pairs are local and lack global coherence, thus not forming a superfluid. This may explain the symmetric lineshape without coherence peaks. Finally, another candidate could be certain hidden order as in the case of URu$_2$Si$_2$ [31]. Further studies are needed to examine these intriguing possibilities.

To our knowledge, our results present the first direct observation of a 2D single layer hydrocarbon molecular crystal that is metallic at elevated temperature and gapped at low temperatures. The weak bonding to the substrate helps to make it an ideal platform to explore novel physics in doped quasi-2D hydrocarbon materials, which have rarely been investigated. Moreover, although the structural and electronic properties of *p*-terphenyl bulk crystals and ultrathin films may differ, our data can shed light on the behaviors of the bulk materials. For example, the nature of the gap observed here is similar to that observed on the K-dosed surface of bulk *p*-terphenyl crystals by photoemission spectroscopy [15].

In summary, we have successfully grown high-quality single-layer PTP/Au films, and systematically studied their structural and electronic properties with different K coverage. Several ordered phases are observed, two of which possess a low energy gap at $E_F$. The gap closes around 50 K without obvious response to magnetic field, and its exact origin is yet to be elucidated.

**Methods:**

The experiments were carried out in a Createc STM system and a Unisoku 1300 STM with magnetic field of 11T. The Au (111) single crystal substrate was prepared by cycles of $Ar^+$ sputtering and annealing at 650℃. *p*-Terphenyl (99%, Alfa Aesar) molecules were then deposited on the Au (111) surface kept at room temperature, using a Knudsen cell. After growth, the sample was cooled in the STM module to 4.5 K and K atoms were subsequently evaporated onto it using SAES alkali metal dispensers. The deposition rate of K atoms was calibrated in advance by directly counting K atoms deposited on Au (111) surface, and was assumed to be constant during the short periods of the experiments. The STM topography was taken in constant current mode, and the dI/dV spectra were collected using a standard lock-in technique with modulation frequency f=943 Hz. A Pt tip was used for all the STM measurements after being treated on a clean Au (111) surface.

**Acknowledgments:**

We thank Prof. J. P. Hu, Prof. N. L. Wang, Prof. X. J. Chen, Prof. Z. B. Huang and Dr. D. C. Peets for helpful discussions. This work is supported by the National Science Foundation of China, National Key R&D Program of the MOST of China (Grant No. 2016YFA0300200) and Science Challenge Project (Grant No. TZ2016004).

**Author contribution:**

M. Q. Ren, W. Chen, Q. Liu, C. Chen and Y. J. Yan performed the film growth and STM/STS measurements and analyzed the data. Y. J. Qiao and Y. J. Chen synthesized the *p*-terphenyl bulk crystals which were used in our earlier experiments. G. Zhou guided the crystal growth and provided useful advice about the lattice structure. T. Zhang, Y. J. Yan and D. L. Feng coordinated the work and wrote the manuscript. All authors discussed the results and the interpretation.

# Supplementary materials

## I. Typical topographic images and dI/dV spectra of as-grown PTP/Au with different Kc

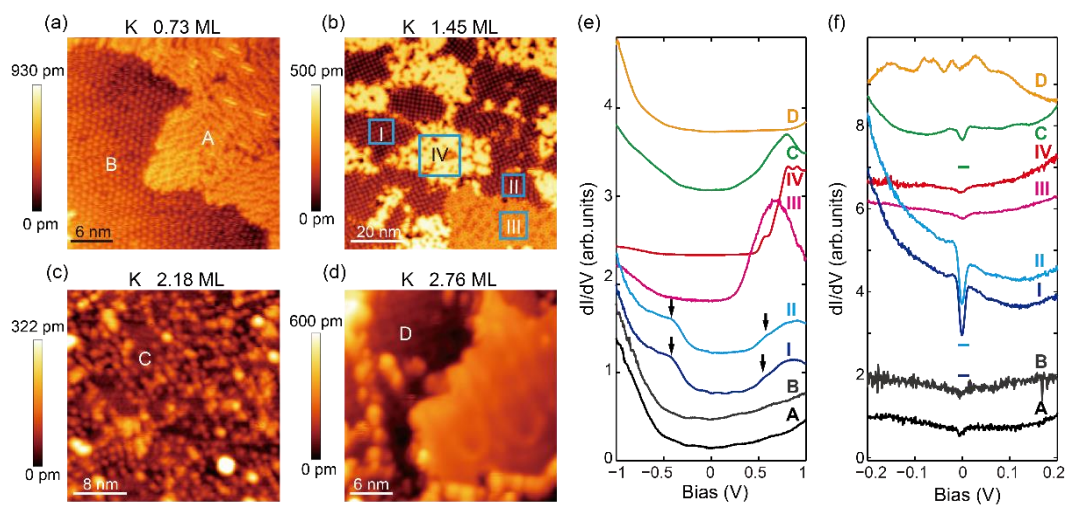

***Fig. s1.*** *Typical topographic images and electronic properties of PTP/Au with different Kc. (a)-(d) Typical topographic images of PTP/Au with different Kc (set point: $V_b$=1 V, I=30 pA). Typical dI/dV spectra of these phases are shown in (e) within the energy range of ±1 V ($V_b$=1 V, I=30 pA, ΔV=30 mV) and (f) within ±200 mV ($V_b$=50 mV, I=100 pA, ΔV=4 mV). Spectra A-D, I-IV are collected in different phases, as marked in (a)-(d). The horizontal bars in (f) indicate the zero conductance positions and colour-coded with corresponding spectra.*

Fig. s1 shows the typical topographic images of PTP/Au with different Kc. For Kc=0.73 ML, two kinds of surfaces are observed, one being a close-packed hexagonal lattice with a lattice constant of 1.0 nm, while the other is much more disordered. By increasing Kc to 1.45 ML, strong phase separation occurs: four typical phases are highlighted by light blue boxes in Fig. s1(b). Phase IV is disordered, while the other three phases possess regular lattice structure (see Fig. 3(b) for phase I, Figs. s2(a)-(c) for phases II, III and IV). These different phases might be induced by different concentrations of K atoms intercalated into the *p*-terphenyl molecules. By further increasing Kc, the surface becomes more and more disordered, until finally only K islands can be observed. The evolution of the electronic structure of PTP/Au with different Kc is shown in Figs. s1(e) and s1(f). Within the energy range of ±1 V, no obvious DOS features are observed in the phases with low Kc, while distinct band structure is visible in phases I and II with Kc=1.45 ML (as indicated by black arrows in Fig. s1(e)). This is further confirmed by the dI/dV spectra within the energy range of ±200 mV, where a surprising gap appears in phases I and II with Kc=1.45 ML.

## II. Typical topographic images and dI/dV spectra of phases II-IV observed in as-grown PTP/Au with Kc=1.45 ML

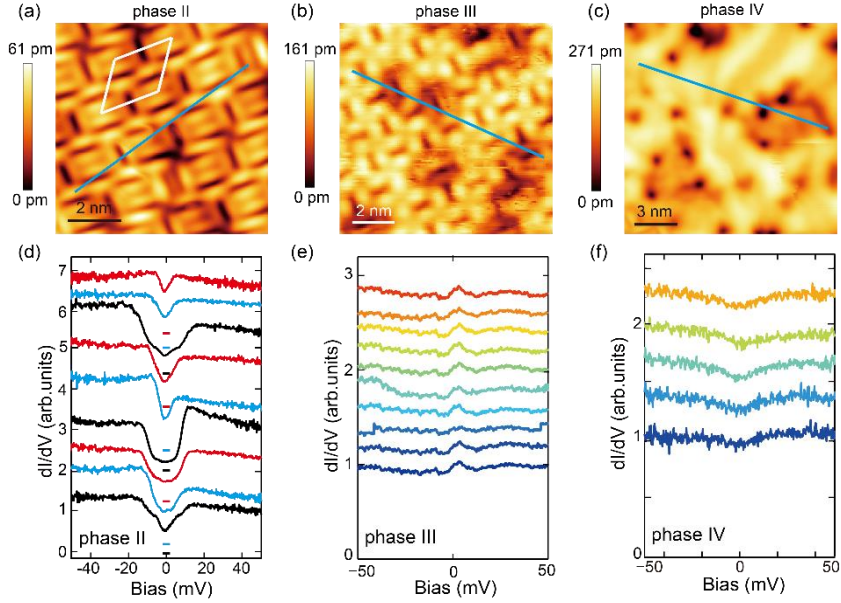

*Fig. s2.* Typical topographic images and dI/dV spectra of phases II-IV observed in as-grown PTP/Au with Kc=1.45 ML. (a)-(c) Typical topographic images of phases II-IV, measured in the areas indicated by light blue boxes in Fig. s1(b) (set point: $V_b$=0.5 V, I=30 pA). Unit cell of phase II is indicated by white box. (d)-(f) Typical low energy dI/dV spectra of phases II-IV, which are collected along the blue lines marked in (a)-(c), respectively (set point: $V_b$=10 mV, I=100 pA, ΔV=1 mV). Curves in (d)-(f) are normalized to their values at +50 mV, and are shifted vertically for clarity. The horizontal bars in (d) indicate the zero conductance positions and colour-coded with corresponding spectra.

Fig. s2 shows typical topographic images and dI/dV spectra of phases II-IV observed in as-grown PTP/Au with Kc=1.45 ML.

(1) Phase II is well ordered, appearing like the alternate stacking of chains of rectangle-shaped blocks and individual *p*-terphenyl molecules. The unit cell is a parallelogram with edges of about 21.2 Å and 18.5 Å and an angle of about 60 °, as indicated by the white box in Fig. s2(a). A gap-like structure similar to that in phase I is also observed in this phase, as shown in Fig. s2(d).

(2) Phase III exhibits a completely different ordered structure, composed of edge-shared windmill-like units. Its low energy dI/dV spectrum is featureless, behaving like a metal (Figs. s2(b) and s2(e)).

(3) Phase IV is disordered, with apparent K clusters distributed randomly on the surface, and also behaves like a metal (Figs. s2(c) and s2(f)).

## III. Detailed topographic images and possible lattice structure of phase I observed in PTP/Au with Kc=1.45 ML

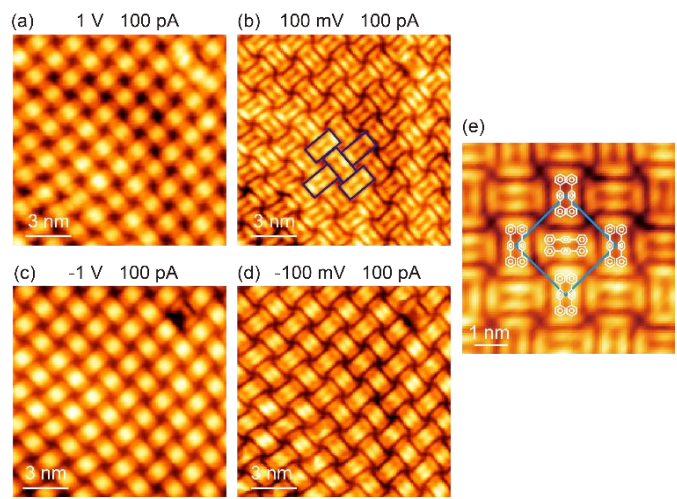

***Fig. s3.*** *(a)-(d) Topographic images of phase I at various sample biases (14 × 14 nm$^2$). (e) Sketch of possible lattice structure.*

To determine the origin of the gap-like feature, it is of vital importance to determine the arrangement of the terphenyl molecules in phase I. Figs. s3(a)-(d) show typical topographic images of phase I taken under different sample biases. Phase I is composed of alternately stacked rectangular blocks along the horizontal and vertical directions, as shown in Fig. s3(b), forming a simple 2D square lattice. The central distance between the nearest rectangular blocks is 1.5 nm, which is very close to the length of *p*-terphenyl molecule. At high sample bias, the centers of the rectangular blocks are much brighter than the edges. On the contrary, the edges of the rectangular blocks are clearly distinguished at low sample bias. Considering that the areal density of the molecules should be the same with or without K doping, we conclude that one such rectangular block consists of two *p*-terphenyl molecules. The chemical bonds within each *p*-terphenyl molecule are very stable, thus K atoms are more likely intercalated into the space between *p*-terphenyl molecules, rather than breaking the original bonds to form a new one. In order to maximize the overlap of the *p*-terphenyl π-bonds and the itinerant electrons from the K atoms, the benzene rings in *p*-terphenyl molecules should be aligned face-to-face.

One lattice model is proposed, as shown in Fig. s3(e). Two *p*-terphenyl molecules lie parallel to each other in the same layer, forming a rectangular block. The bicephalous benzene rings lie flat on the substrate while the middle benzene rings may stand up. K atoms are intercalated in the space between the two middle benzene rings. This model can explain the topographic images at different sample biases very well, but theoretical simulations are needed to verify this lattice structure.

## IV. Temperature dependence of topographic images of phase I observed in as-grown PTP/Au with Kc=1.45 ML

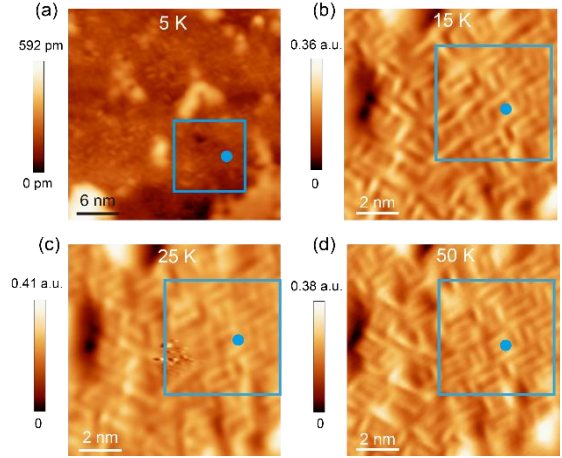

*Fig. s4.* Temperature dependence of the topography of phase I observed in as-grown PTP/Au with Kc=1.45 ML. (a): real space image, (b)-(d): differential images for clarity. The measured area is kept the same at different temperatures, as indicated by the blue boxes, and its topography remains the same at temperatures up to 50 K. U-shaped gaps exist in half of the ordered areas. Temperature dependence of a typical U-shaped gap shown in Fig. 3(e) was collected at the same location, as marked by blue dots.

## V. Impurity state observed in as-grown PTP/Au with Kc=1.45 ML

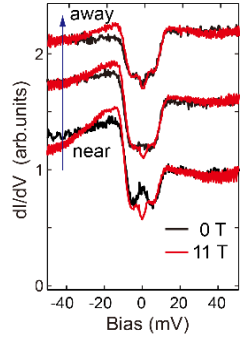

*Fig. s5*. Magnetic field dependence of a U-shaped gap measured around an impurity. Spectra measured at different locations are shifted vertically for clarity.

As mentioned in the main text, a smaller gap is commonly observed inside the U-shaped gap under H=11 T, but is absent at some locations under H=0 T, as shown in Fig. 4(f). Here we clarify that the absence of the smaller gap at some locations under H=0 T is induced by impurity effect. As shown by the black curves in Fig. s5, a strong peak appears around $E_F$ near an impurity, and gradually disappears away from the impurity, and finally a smaller gap appears. Under H=11 T, the peak around $E_F$ is suppressed completely, and the smaller gap appears ubiquitously inside the U-shaped gap. Such impurity state can also be resolved in ZBC maps shown in Figs. 4(c) and 4(d).

Some bright points (yellow dots in Fig. 4(c)) are observed in ZBC map under H=0 T, but disappear in the ZBC map under H=11 T.

## VI. Response of superconducting gap to magnetic field in 1ML FeSe/STO

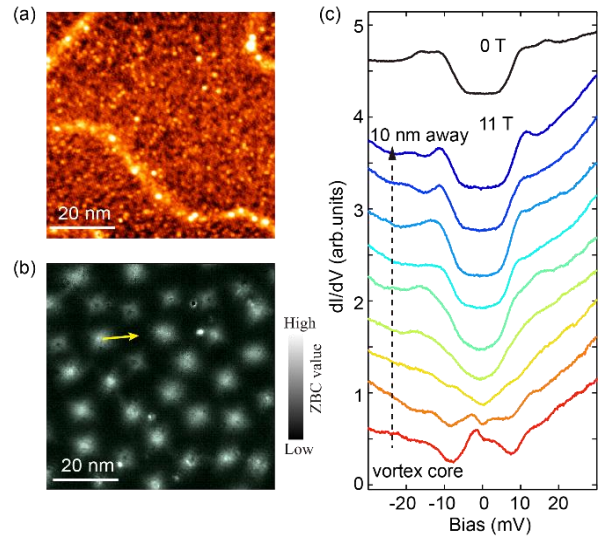

***Fig. s6***. *Magnetic vortex states of single-layer FeSe/SrTiO$_3$ (001) [27]. (a) Topographic image of the surface area for dI/dV mapping (size: 70 × 70 nm$^2$). (b) Zero-bias conductance (ZBC) map taken in the area shown in (a) under H= 11 T (set point: V$_b$ = 30 mV, I = 100 pA, ΔV= 1 mV) shows the emergence of the vortex lattice. (c) A series of spectra taken along the yellow arrow in (b) (set point: V$_b$ = 30 mV, I = 60 pA, ΔV = 1 mV). The uppermost spectrum is taken under H = 0 T, included here for comparison. The superconducting gap is completely suppressed in the core center, and gradually recovers away from the vortex core. The superconducting spectra 2ξ away from the vortex core (ξ is 2-3 nm in 1 ML FeSe/STO) are basically the same as those measured under H =0 T. See Ref. 27 for more details.*